\documentclass[a4paper,conference]{IEEEtran} 
\ifCLASSINFOpdf
  \usepackage[pdftex,dvipdfm]{graphicx}
\else
\fi
\usepackage[cmex10]{amsmath}
\usepackage{amssymb,verbatim}
\usepackage[dvipdfm]{graphicx}
\begin{document}
%
\title{%
Quantum stream cipher $\&$ Quantum block cipher \\
-The Era of 100 Gbit/sec real-time encryption-}

\author{
\IEEEauthorblockN{Osamu HIROTA$^{1,2}$ \\}
\IEEEauthorblockA{
1. Quantum ICT Research Institute, Tamagawa University\\
6-1-1, Tamagawa-gakuen, Machida, Tokyo 194-8610, Japan\\
2. Research and Development Initiative, Chuo University, \\
1-13-27, Kasuga, Bunkyou-ku, Tokyo 112-8551, Japan\\
{\footnotesize\tt E-mail:hirota@lab.tamagawa.ac.jp} \vspace*{-2.64ex}}
}

\maketitle

\begin{abstract}
This paper is the part-II of the previous paper and introduces the world of Yuen's concept.
In the theory of cryptology, the Shannon impossibility theorem states that the upper bound of 
the security of a plaintext against 
a ciphertext-only attack is the entropy of the secret key. 
At the same time, it gives the upper bound of the unicity distance against a known plaintext attack.
Hence the development of a new symmetric key cipher requires finding a way to undo or lift this theorem.
Such challenges have been attempted with quantum stream cipher and quantum data locking as block cipher.
The both ciphers are designed by means of differentiating the receiving performance of 
Bob with key and Eve without key according to the principle of quantum communication theory. 
Thus, the origin of security of both ciphers come from the principle of keyed communication in 
quantum noise (KCQ) proposed by Yuen.
In this paper, we explain and compare the principles and features of both ciphers, and assist to 
improve the quantum data locking scheme.
Then we will introduce experimental research on quantum stream cipher towards commercialization, 
which has performance superior to conventional cipher.

\end{abstract}

%
\IEEEpeerreviewmaketitle

\section{\textbf{Introduction}}
Shannon's founding of modern communication theory was also the founding of cryptography based on it. 
However, contrary to Shannon's intention, his theory was named information theory, which caused great confusion 
in its subsequent development. It is well known that this arose from a misunderstanding of the operational meaning of Shannon entropy. 
However, researchers in information science and cryptology overcame this difficulty and brought about the revolution known 
as coding theory and computational security for the Global Information and Communication Technology.
However, the same confusion is not resolved in quantum information science.
Yuen has repeatedly explained the seriousness of this situation, but the effect appears to be small.
His argument was based on the history of the development of quantum communication theory at MIT as follows.

In order to further develop the general Shannon communication theory including quantum communication theory developed by pioneers such as 
Helstrom, Holevo, and Yuen, toward contributing to a real world society, a joint USA-USSR workshop was held in 1975 (Fig.1). 
On the subject on quantum communication theory, R.S.Kennedy of MIT played a central role with the support of Helstrom.
In addition, T. Berger, R. Gray, T. Cover and others from the field of  American information theory participated, 
and the importance of clarifying operational meaning in the formulation of the mathematical generalization of information theory 
was discussed.
The NATO workshop in 1986 was important in terms of further technical developments of quantum technologies (Fig.2).
Given this history, Yuen's caution is natural. But confusion continues in the field of quantum cryptography.

\begin{figure}
\centering{\includegraphics[width=6cm]{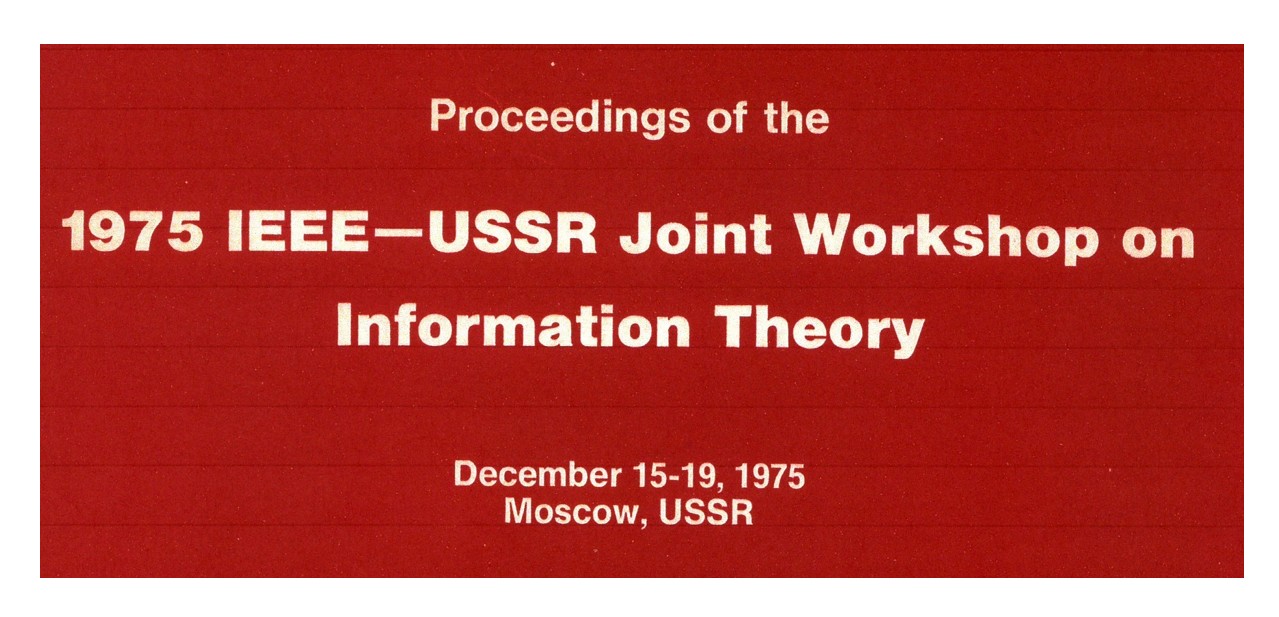}}
\caption{This workshop was held under the agreement on cooperation in information theory between V.I.Siforov (USSR) and J.K.Wolf (USA)
The chairman was G.D.Forney Jr, and the number of contributors is 29. Kennedy introduced the results of Helstrom, Yuen, Dolinar, 
Davis Holevo, Belavkin, Stratonovich, Bakut, and Shchurov. }
\end{figure}

\begin{figure}
\centering{\includegraphics[width=7cm]{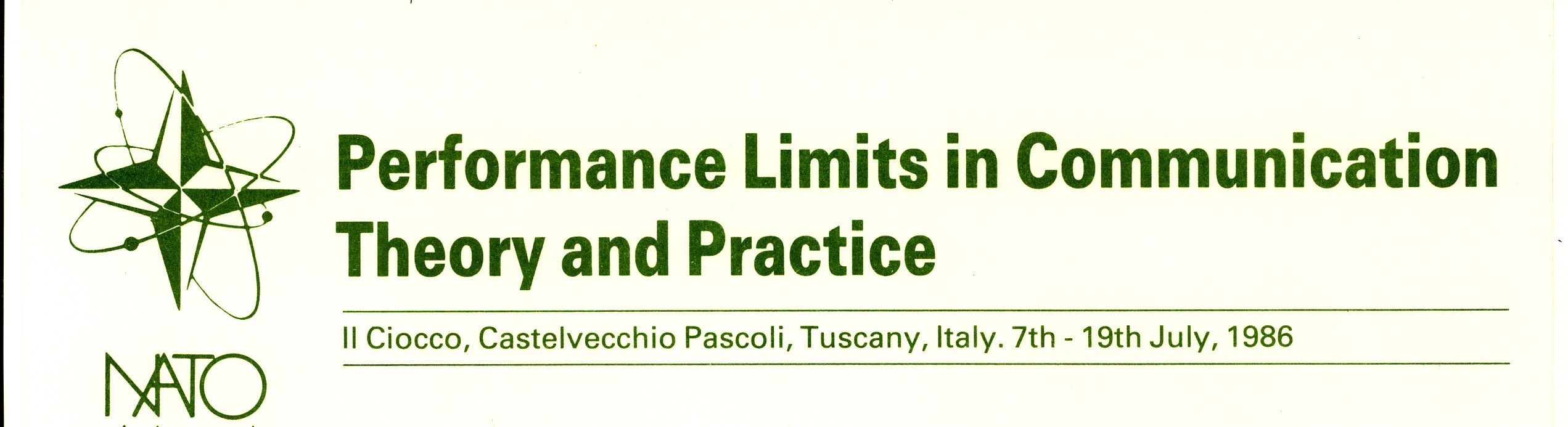}}
\caption{When discussing the limits of communication theory, the operational meaning of the evaluation function is 
of utmost importance. 
All evaluation functions are defined with technical meanings that lead to their usefulness to society. 
Current Internet technology has matured from discussions at such conferences. 
Therefore, quantum information science also requires careful consideration. 
Chairmans for sessions of physical limitation and optical communication were L.B.Levitin, and O.Hirota, respectively. }
\end{figure}

Under such a situation, in 2000, Yuen disclosed the framework for quantum stream cipher which is a new symmetric key 
encryption concept that aims to circumvent Shannon's impossibility theorem.
This is a technology to protect data by differentiating signal reception capabilities based on 
quantum communications theory. It is not a concept that has been used in the past of hiding data by devising new encryption protocols.
  
Roughly speaking, this new concept can be said to be a prescription for retaining the convenience of mathematical encryption
 while enhancing the security of that encryption capability through the use of quantum noise.
Therefore, the communication system or modulation scheme itself becomes the encryption function.
As a result, an eavesdropper is battling against physical ciphertext in which a mathematically encrypted ciphertext is hidden by noise.
We have previously introduced this new cryptographic concept in three papers with the same title [1,2,3]. 

To further develop them, 
in the section II and III we explain the basic concept of the background of the quantum stream cipher and quantum data locking. 
The section IV discusses the origin of lifting the Shannon impossibility theorem.
In the section V, we propose a general quantum block cipher. The section VI introduces an example of the general quantum block cipher.
The section VII introduces some experimental progresses of the quantum stream cipher currently in development.

\section{\textbf{Concept of generalized random cipher and quantum symmetric key ciphers}}
From 1998 to 1999, Yuen discussed the concept of generalized random cipher in our seminars. 
He advocated the construction of a theory of a large framework including QKD. 
I suggested that he should discuss symmetric key cipher separately from the general theory, 
because the direction of the theory could confuse many researchers. 
This section provides a brief explanation of his concepts on the quantum symmetric key cipher.

\subsection{\textbf{Operational meaning of criteria}}

The operational meaning of a criterion in technology theory is a concept established for 
the quantitative comparison of technical effectiveness.

Originally, information-theoretic security means the inability of a cipher to be decrypted except by probabilistic inference. 
It never claimed that it was impossible to decrypt.
In the quantum field, the interpretation is widespread that it means ciphers that are absolutely unbreakable. This is incorrect.

When designing an information-theoretically secure cipher, it is necessary to define the quantitative characteristics of 
its security and make sure that it has operational meaning.
Theories for evaluating information-theoretic security were discussed by Shannon-Massey et al. It has a long history.
These are unified as the theory on unicity distance and the spurious key based on the conditional entropy.
\begin{equation}
H(X|Y)=\sum_{y\in Y} p(Y)\sum_{x\in X}p(X|Y)\log \frac{1}{p(X|Y)} > 0
\end{equation}
Then, its operational meaning is very clear for the application to the technical field [4,5].

On the other hand, one has the similar creiteria so called the Shannon mutual information as follows:
\begin{equation}
I(X;Y)=H(X)-H(X|Y)=H(Y)-H(Y|X)
\end{equation}
This does not mean ``information of data" or ``number of bit of signal sequence", but the operational meaning is 
the efficiency of coding and decoding of target communication channel.
Interpretations such as the correlation between A and B are mathematical relationships and do not imply operational meaning.
A lot of researchers really misundersand this in the quantum information science. The situation is very similar to 
the 1970s $\sim$ 1980s as denoted in the introduction..

Thus, following the conventiional theory, we have adopted guessing probability (or error probability) and 
the unicity distance theory based on conditional entropy to evaluate the generalized random ciphers. 
The detailed discussions on the generalized unicity distances for such a cryptographic mechanism have been given in  [2,3].
The following is the short summary for those discussion.

\subsection{\textbf{Generalized random cipher}}
There are several definitions for the concept of randomization. Initially, it was related to the concept of a typical sequence 
by Shannon.
But we deal with cipher which has an operation called randomization.
We call it Shannon-Massey random cipher [4,5].
In this definition, so far private randomization of plaintext is considered to be the representative of random encryption. 
These aim to strengthen the information-theoretic security of conventional cipher.
However, due to the Shannon impossibility theorem explained below, there are limitations to their performance.
In response to this, Yuen defined the technology system that randomizes ciphertext.
This section provides a brief explanation of this.

In general, the security of ciphertext only attack on data is bounded by the following theorem [4,5].\\

\textbf{Theorem-1} (Shannon impossibility theorem)\\
In the ciphertext only attack on data, the security is bounded by the entropy of the secret key as follows:
\begin{equation}
H(X|Y) \le H(K)
\end{equation}
In addition, it makes it more difficult to guarantee security against known-plaintext attacks on the key.

In order to undo the above theorem, Yuen showed the following theoretical situation [1,2,3,6,7] (Fig.3):

\textbf{Theorem-2}\\
The necessary conditions for lifting the Shannon impossibility theorem are:\\
\textbf{(a)} The ciphertext received by a legitimate receiver (Bob) and the ciphertext received by 
an eavesdropper(Eve) are different.$Y^B_n \ne Y_n^{E}$\\
\textbf{(b)} The following properties are required. 
\begin{eqnarray}
H(Y^B_n|K,X_n)&=&0 \\
H(Y^{E}_n|K,X_n) &\ne& 0
\end{eqnarray}
where $Y^B_n,Y_n^{E}$ are ciphertext for Bob and Eve,and $X_n$ is plaintext. $K$ means secret key.\\

Such ciphers are called the generalized random cipher. 
If such a situation could be implemented, it would be possible to realize a symmetric key cipher 
that has information-theoretically secure against known plaintext attacks even using a PRNG with a short key.
Furthermore, the possibility appears such that Eve cannot obtain the correct plaintext even with the correct key after communication. 
These features mean that the Shannon impossibility theorem in the cryptology can be undone. 

\begin{figure}
\centering{\includegraphics[width=7cm]{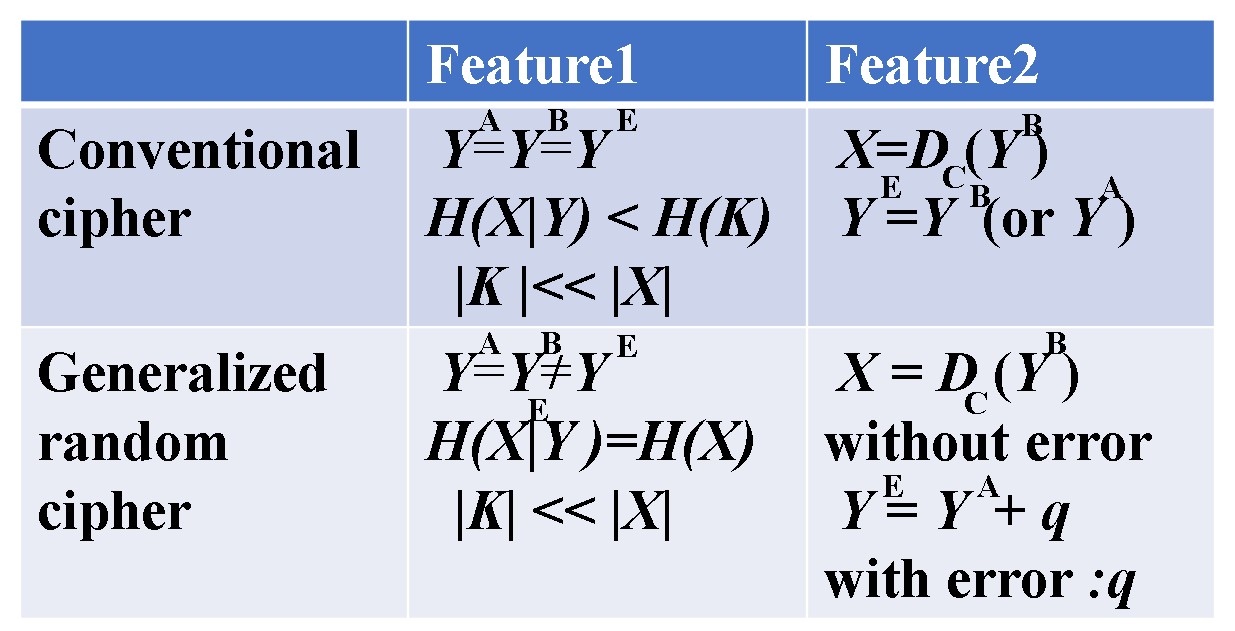}}
\caption{The conceptual differences between conventional mathematical cipher and generalized random cipher. 
$X$ is plaintext as natural sentence, $Y$ is ciphertext, $K$ is shared key, $|K|$ is key length, $D_C$ is decryption. 
A, B, E are indexes for Alice, Bob, and Eve.
Here, the following abbreviations are used:$Y^{E_q}=Y^E$ which is Eve's ciphertext with error. }
\end{figure}

\begin{figure}
\centering{\includegraphics[width=8cm]{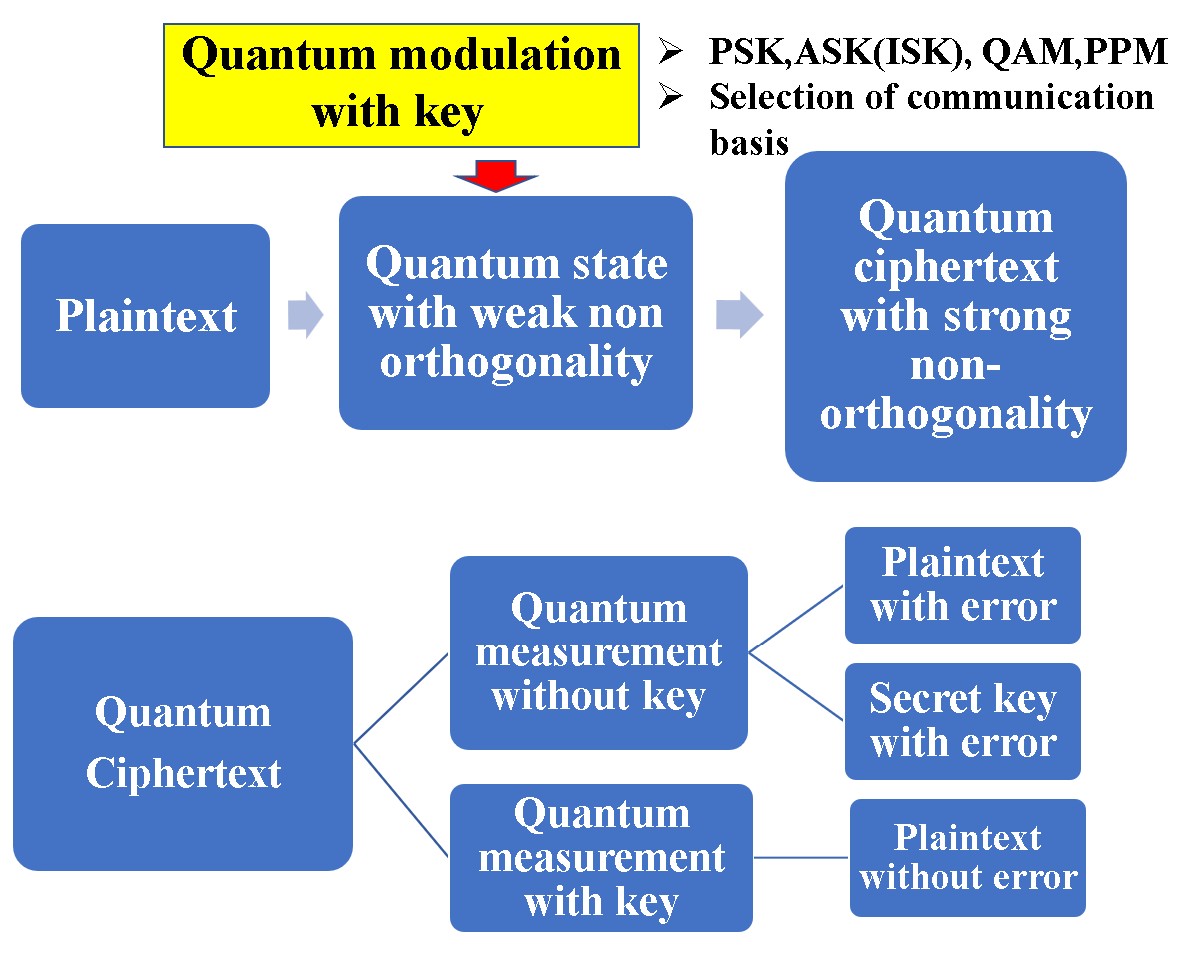}}
\caption{An example of KCQ (keyed communication in quantum noise). The figure above shows an encryption process, 
and the figure below shows a decryption process. The key of Alice is used to increase the non-orthogonality 
as the ensemble of quantum states in the sense of quantum communication theory, and the key of Bob is used to 
weaken the non-orthogonality. For example, binary (with key) vs multi-ary quantum states (without key.}
\end{figure}

\subsection{\textbf{Keyed communication in quantum noise: KCQ}}

In order to realize the generalized random cipher, Yuen provided the concept of KCQ (Keyed communication in quantum noise) 
in 2000 (Fig.4).
He uses the following theorem in the quantum communication theory [8,9,10].\\

\textbf{Theorem-3} (Helstrom$\cdot$Holevo$\cdot$Yuen)\\
A set of non orthogonal quantum states cannot be identified without error. 
The bound of the error is given by the quantum detection formulation.\\

We pick up a simple example from Yuen's concept on the generalized random cipher.
Here, let us assume that  binary classical data is transmitted by a communication basis $\{|\psi_1>, |\psi_2>\}$ 
which is composed by non-orthogonal states with characteristics close to orthogonal.
Large number of communication basis are prepared.
One communication basis is randomly selected using a secret key (including PRNG), and binary information is transmitted using 
the selected communication basis. 

If quantum states in the set of communication basis have strong non-orthogonality, 
then a large difference in error characteristics will occur between a receiver with the secret key and a receiver without 
the secret key.

In 2000, Yuen and Kumar showed a concrete system model of the generalized random cipher 
(code name is $\alpha\eta$) based on KCQ principle. This is at present called the standard quantum stream cipher.
The first experimental demonstration was reported in QCMC 2002 by Kumar group [11].

\section{\textbf{Evaluation of security performance}}

\subsection{\textbf{Quantum symmetric key cipher}}

When plaintext is mapped to a quantum state, the set of quantum states is called a quantum ciphertext. 
The probability of guessing the plaintext or secret key by quantum measurement of the quantum ciphertext is the foundation of security. 
The theory that quantifies it is quantum signal detection theory, which is the core of quantum communication theory.
That is, the criterion for evaluating the information-theoretic security of symmetric key cipher is in general 
the guessing probability for data, and 
the unicity distance or spurious key for secret key which is defined by conditional entropy derived from 
the guessing probability analysis. 
In addition, these must be premised on guessing probabilities that are based on 
accurate error probabilities of transmission signals.
This has been used for a long time in cryptography because it has a very clear operational meaning. 

Given this historical background, the following security formulations for the generalized random cipher were 
defined to evaluate the characteristics of the generalized random cipher [1,2,3]:

\textbf{(a)} \textbf{Ciphertext only attack on data} 

The measurement of the signal corresponding to the plaintext by Bob, who has the key, is almost error-free, 
while the measurement by Eve, who does not have the key, has a large error. 
We use  the conditional entropy based on the guessing probability $p(X|Y)$ derived from the analysis of the error probability 
($p(Y|X)=Tr \rho\Pi$) which is a major theme in quantum detection theory.
The problem is to check whether it is possible to:
\begin{equation}
 H(X^E_n|Y^E_n) \le H(K) \longrightarrow  H(X^E_n|Y^E_n) > H(K)
\end{equation}
based on the error performance of quantum mesurement.

\textbf{(b)} \textbf{Ciphertext only attack on key} \\
The unicity distance of a ciphertext only attack on key is defined for the eavesdropper's ciphertext as follows:

{\bf Definition-1}: \\
Let $n^Q_0$ be the minimum length of the ciphertext that has zero key ambiguity for the eavesdropper's ciphertext.
Then it is given by
\begin{equation}
n^Q_0: H(K|Y^{E}_{n^Q_0})=0
\end{equation}
 $n^Q_0$ is called the unicity distance of ciphertext only attack for generalized random cipher.

Unlike the conventional type, the above equation does not depend on the statistical structure of the plaintext, 
but on the randomness of the ciphertext that can be obtained by the eavesdropper.

\textbf{(c)} \textbf{Known plaintext attack on key}\\
The conventional random ciphers can achieve a large unicity distance for a ciphertext-only attack.
However, it is impossible to guarantee information-theoretic security more than a key length in the known plaintext attack.

Here, we show the unicity distance for a known-plaintext attack on key of generalized random cipher.

{\bf Definition-2}:\\
The unicity distance of known plaintext attacks for generalized random cipher is defined as follows:
\begin{equation}
n^Q_1: H(K|X_{n^Q_1}, Y^{E}_{n^Q_1})=0
\end{equation}

In the case of generalized random cipher, at least, the following can be expected.
\begin{equation}
|K| \ll n^Q_1 \le 2^{|K|}
\end{equation}
where $|K|$ is the length of key with a true randomness. This performance is the most important in the practical applications.
This is not possible with existing cryptography theory.

Quantitative evaluation of the KPA security of quantum symmetric key cipher can be done directly by 
the error properties of the eavesdropper's ciphertext. However, for consistency with Shannon theory, it is necessary 
to derive the unicity distance. To evaluate this, we have [6].\\

\textbf{Theorem-4} $\{Yuen \cdot Nair\}$\\
The lower bound of the generalized unicity distance for KPA  is given as follows:
\begin{equation}
n^Q_1 \ge \frac{H(K)}{C_1},\quad C_1=\max_{\{\Pi^E\}}I(K^R;Y^{E_q})
\end{equation}
where  $C_1$ is the maximum mutual information of the channel constructed by quantum measurement of 
quantum states related to running keys $K^R$ from PRNG. \\

This theorem is derived from the correct operational meaning of the mutual information being the encoding-decoding efficiency.
The theory of maximaization of mutual information is twinned with the quantum detection theory and 
has the same theoretical structure as the optimal theory.
That is, according to the Helstrom-Holevo-Yuen formulation, the mutual information of the channel consisted of a set of 
non-orthogonal quantum states can be controlled by the degree of non-orthogonality. 
In fact, the following theorem gives such a situation [9].\\

$\textbf {Theorem-5}$ \{Holevo\}\\
The necessary condition for maximum mutual information with respect to the decision operators is:
\begin{eqnarray}
&&P(j|i) =Tr \rho_i \Pi_j \nonumber  \\
&& {\bf F}_j = \sum_{l} \xi_k \rho_k \log  \{\frac {P(j|l)}{\sum_k \xi_k P(j|k)}\} \nonumber \\
&& \Pi_j[{\bf F}_j - {\bf F}_i]\Pi_i =0,  \forall i,j 
\end{eqnarray}
where $\{\Pi_j\}$ is the quantum decision operaotr.\\

From the above, we can know that the a mutual information for the non-orthogonal states is smaller than that of the orthogonal state. 
Therefore, the goal of this method is to minimize the mutual information 
of the receiver who does not have the key.

This is intended to indicate the lower bound of the unicity distance defined by conditional entropy, 
taking into account the operational meaning of Shannon mutual information [2]. 
The mutual information itself is not directly used to evaluate the security of a cryptosystem.

\subsection{\textbf{Criteria of quantum stream cipher}}

The quantum symmetric key cipher of stream cipher form is called quantum stream cipher. The evaluation has already been discussed 
using unicity distance theory. 
In that evaluation, ciphertext-only attacks on data and key, and known plaintext attacks are the subjects of the evaluation. 
The details have already been discussed. So please refer to the previous section and the literature[1,2,3].

\subsection{\textbf{Criteria of quantum data locking as block cipher}}
In 2004, D.P.DiVincenzo and his group reported the concept of quantum data locking [12]. 
The basic concept is to consider the difference between the Shannon mutual information
(accessible information of the quantum measurement channel) of a receiver with a shared key and 
a receiver without a shared key as follows:

\begin{equation} I^B_{acc}(with ``key") -I^E_{acc}(without ``key") \end{equation}

Here, $I^B_{acc}$ is the mutual information (accessible information) of the keyed A-B channel, and 
$I^E_{acc}$ is the mutual information (accessible information) of the unkeyed A-E channel. 
They interpret this difference as ``the amount of information" or ``classical correlation" being locked.
However, the operational meaning of Shannon mutual information is the efficiency of the coding of the communication channel. 
It is not the information amount of the data. As a result, discussions on cryptographic performance, 
such as quantitative evaluations of known plaintext attacks, are difficult. 

Therefore, evaluations are made through indirect discussions. Even if they adopt the variational distance as follows [13]; 
\begin{equation} \sum_{X}|p(X) - p(X|Y)| \le \epsilon \end{equation}
the situation does not improve. We resolve these in Section V.

\section{\textbf{Origin of lifting the Shannon impossibility theorem}}
A necessary condition for dissolving the Shannon impossibility theorem is that the ciphertexts received by Bob and Eve 
are different, according to Yuen. Here we will discuss how this can be achieved by limiting ourselves 
to ciphertext-only attack on data (plaintext).\\

\subsubsection{\textbf{Quantum stream cipher}}
We discuss a reason of why the Shannon impossibility theorem in the quantum stream cipher is undone.  
Let us use a simplified model of quantum stream cipher form as follows [1,2,3,6].\\ 

\textbf{(a) Encryption} 

A pair of two coherent states that transmit binary classical data $X$ are used as communication basis
\begin{equation}
{\bar \theta} = \{|\alpha >, |\alpha e^{i\pi}>\}, \quad X=0,1
\end{equation} 
and many communication bases with different classical parameters $\{{\bar \theta}_j\}$ are prepared. 
Then, one communication basis from a set of 
\begin{equation}
\{{\bar \theta}_j\}=\{|\alpha e^{i\theta_j}>, |\alpha e^{i(\theta_j +\pi)}>\}:  j=1,2,\dots, M
\end{equation}
is randomly selected using running key $j \in K_R$ from PRNG with a true random secret key :$K$. 
Here, let the operation of selecting a communication basis be the unitary map $U_{K_R}$.
This corresponds to phyisical encryption. 
Here the operation of the unitary map is as follows:
\begin{equation}
U_{K_R}:\{|\alpha>,|\alpha e^{i\pi}>\} \mapsto  \{|\alpha e^{i\theta_j}>, |\alpha e^{i(\theta_j +\pi)}>\}
\end{equation}
The running key ($K_R$) corresponds to $j$.
Then classical binary data is transmitted using selected communication basis. 
As a result, one of $2M$ quantum coherent states is transmitted on the channel in each time slot.
The set of $2M$ coherent states corresponds the quantum ciphertext.\\

\textbf{(b) Decryption} 

Bob knows the key, it can perform an inverse unitary map $U^{-1}_{K_R}$ 
with shared information. The inverse map transforms all bases into the starting 
communication basis as follows: 

\begin{equation}
U^{-1}_{K_R}:\{|\alpha e^{i\theta_j}>, |\alpha e^{i(\theta_j +\pi)}>\}\mapsto 
\{|\alpha>,|\alpha e^{i\pi}>\}, \forall j
\end{equation}

Therefore Bob always can receive binary PSK form. As a result, the error can be made extremely small. 
Since Eve does not know the secret key, she has to assume that the signal (quantum ciphertext) of composite information of 
data and key is transmitted by using one of all possible communication basis.  
It will result in a large error, because of Helstrom-Holevo-Yuen formulation.

In other words, there will be a difference in the reception characteristics when the key is known and when it is not.
This is derived from the principle of quantum communication theory that multi-ary signals have stronger non-orthogonality 
than binary signals, resulting in deterioration of discrimination performance.
This is called ADVANTAGE CREATION by the secret key.\\

\textbf{(c) Eavesdropping} 

\textbf{(i)} When Eve considers a signal system as being in $2M$-valued quantum state and infer binary data 
from its reception, her problem becomes to distinguish adjacent signals $(|\alpha e^{i\theta_j}>, |\alpha e^{i\theta_{j+1}}>)$ 
in $2M$-valued signal, where $\theta_{j+1}=\theta_j + \pi/M$. This is because data 0 and 1 are set alternately in the $2M$ phase signals.
Two signals are completely masked by quantum noise, so ${\bar {P}_e^E}=1/2$.

\textbf{(ii)} On the other hand, from the structure of encryption, Eve can try to attempt to decode the binary data directly.
At that time, she would adopt the binary quantum optimal measurement for the following mixed quantum states at each time slot [14].
\begin{eqnarray}
\rho^E_0 &=&\frac{1}{M}\sum_{m=1}^{M} |\alpha_{(m=even)}><\alpha_{(m=even)}| \nonumber \\
\rho^E_1 &=&\frac{1}{M}\sum_{m=1}^{M} |\alpha_{(m=odd)}><\alpha_{(m=odd)}| 
\end{eqnarray}
This structure of mixed state is called doubly symmetric mixed state, and the quantum Bayes(also minimax) solution for the above states
 was given in [14].
Then the average error probability at each slot for binary data under the condition: $M \gg 1$ is given as follows: 
\begin{equation}
{\bar {P}_e^E}=\max_{\{\xi\}}\min_{\{\Pi\}}\{1-\frac{1}{2}\sum_{l=0}^{1} Tr \rho_l^E \Pi_l \}\longrightarrow \frac{1}{2}
\end{equation}

In addition, one can enjoy some additional randamizations based on mapping modulation such as OSK (Overlap selection keying) [1,2,3,15].
As a result, we have always
\begin{equation}
\rho^E_0 = \rho^E_1 \longrightarrow \bar{P}^E = \frac{1}{2}
\end{equation}

Thus, it is impossible to discriminate such states at each time slot, and Eve's measurement result is independent of any plaintext. 
Here we simplify the symbols.
Then, we have the following relation for any plaintext sequence $X_n$.
\begin{equation}
H(X^E_n|Y^E_n)=H(X^E_n) > H(K),  under |K| \ll |X_n|
\end{equation}
 Here $X_n$ and $K$ are the plaintext and secret key of Alice, and 
$X^E_n$ and $Y^E_n$ are the plaintext and ciphertext that Eve can obtain.

\begin{figure}
\centering{\includegraphics[width=8cm]{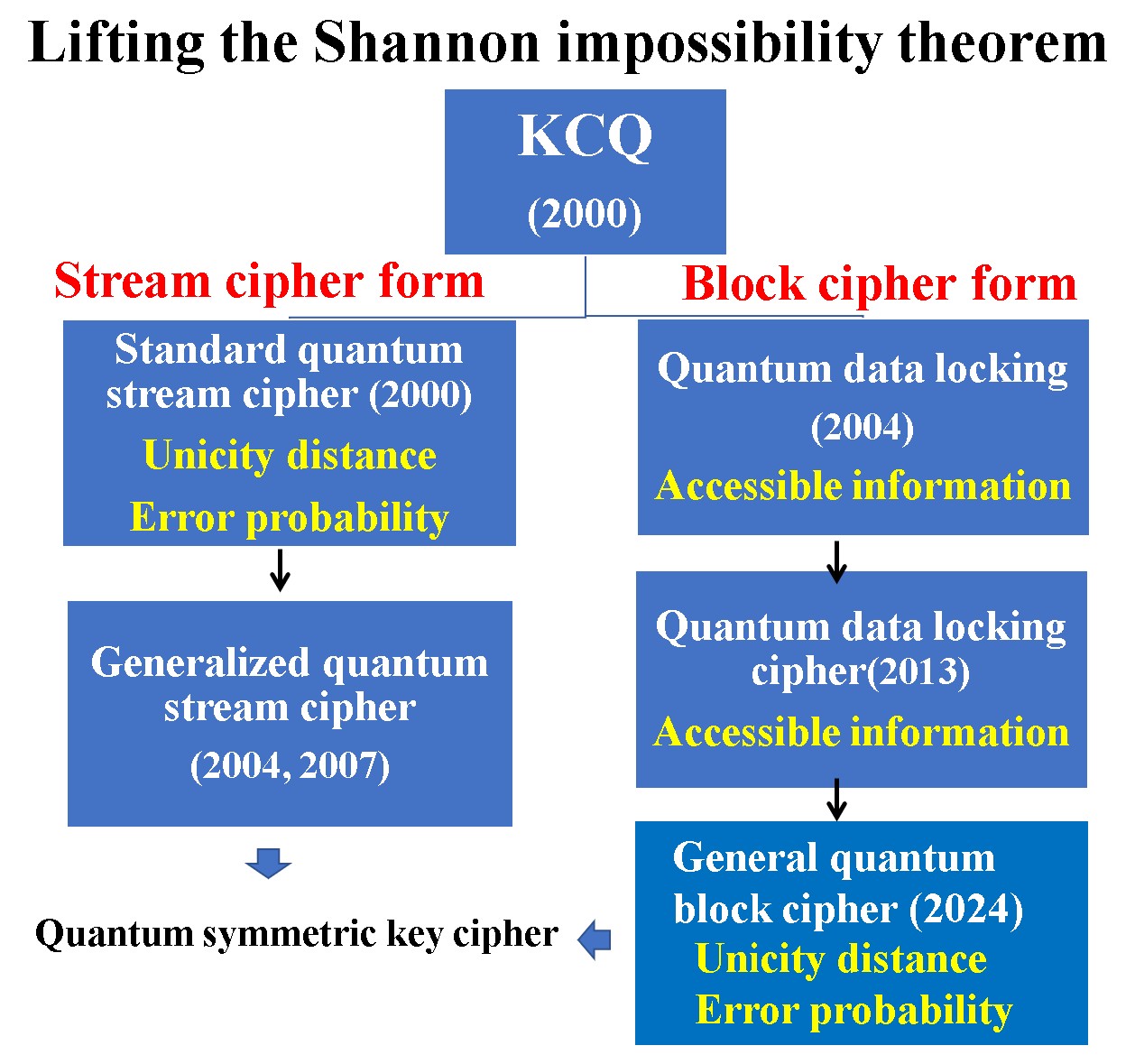}}
\caption{The origin of the lifting the Shannon impossibility theorem is KCQ (Keyed communication in quantum noise). 
As a method for achieving it, standard quantum stream cipher was proposed in 2000 and experimentally demonstrated in 2002. 
Generalized version was given in 2004and 2007.
In 2004, a method called quantum data locking was proposed from a different perspective of KCQ, 
and in 2013 it began to be discussed as an encryption method. 
Security criteria is unicity distance and accessible information, respectively. }
\end{figure}

However, since pseudorandom numbers are used, it becomes necessary to additionally ensure information-theoretic security 
for the secret key $K$.
To ensure sufficient information-theoretic security for the key, the above scheme needs to be generalized.
See the references [1,2,3] in which the delaied explanation is given.\\

\subsubsection{\textbf{Quantum data locking as block cipher}}

They adopt the following implementation scheme.

\textbf{(a)} The sender Alice maps classical information $|X|=2^n$ of $n$ bits into an orthogonal quantum state, 
and then uses a unitary map: $U_k, k\in K$  to project it into a set of non-orthogonal quantum states. $K$ is the true random number. 
So the selection of the unitary map is equivalent to Haar measure.\\

\textbf{(b)} The receiver with the key converts the set of non-orthogonal state signal back into 
a set of orthogonal state as a basis and receives the information without error by means of the inverse unitary map :$U^{-1}_k$. 
This means that the total information from the sender can be obtained.
A receiver without the key can only obtain information of signals via the set of non-orthogonal states. \\
Hence one can know that the concept of this type of cipher is the differentiation by means of with key and without key.
Furthermore, this differentiation arises from the fact that a non-keyed receiver needs to distinguish between 
sets of quantum states with strong non-orthogonality.

Thus it is  another way of expressing Helstrom-Holevo-Yuen formulation of the indistinguishability of non-orthogonal states.
The settings up to this point are the same as KCQ (Fig.5).

Here we discuss what is different between the quantum stream cipher and the quantum data locking.
From (a) and (b), this system corresponds to a block cipher form that encrypts $n$-bit block plaintext by the secret key of $|K|$ bits.
They interpret the difference in mutual information (Eq(12)) as the number of bits that can be securely encrypted with 
a $|K|$-bit secret key.
That is, they use to evaluate the cryptographic feature the following function [12,13].
\begin{eqnarray}
\eta&=&\frac{H(K)}{I_{acc}(with``key") -I_{acc}(without``key")}\nonumber \\
&=& \frac{H(K)}{H(X|Y)}
\end{eqnarray}
Consequently, if the situation of $\eta < 1$ is possible, it provides the following relation like Eq(21):
\begin{equation}
H(X|Y)=\frac{1}{\eta} H(K) > H(K), under |K| \ll |X|
\end{equation}

We can see an example. 
Let us here require the following upper bound:
\begin{equation}
I_{acc}(without``key") < \epsilon \log |X|,\quad 0 < \epsilon < 1
\end{equation}
then Fawzi et al showed [13] that the required secret key entropy is : 
\begin{equation}
H(K) \cong 4\log \frac{1}{\epsilon}
\end{equation}
In this case, by making $n=\log |X|$ larger, $\eta$ can be made arbitrarily small.
For example, one has the following[15].
\begin{equation}
\eta \sim \frac{\log n}{n} \ll 1, \quad n \gg 1
\end{equation}
Thus they claimed that the Shannon impossibility theorem is violated.
This is coincidence in the special case of the operational meaning of mutual information.

On the other hand, quantum stream ciphers use the same mutual information, but do use its correct operational meaning
 (efficiency of coding and decoding) to evaluate security in the unicity distance theory under 
the analysis of the error probability of signals.

In addition, the formalism of quantum data locking does not include any issues that are absolutely necessary for 
encrypted communication for high speed data flow, such as a theoretical construction on the subject of time axis 
and its associated key requirements, communication rate (bit/sec efficiency) and bandwidth requirement, 
delay at processing and so on. 
Especially user requires that it handles at least real time long data of 10Gbit/sec.
Thus, further refinements are needed to apply quantum data locking to practical communications. 

In order to put these ideas into practical use, we will provide methods to solve the above problems, converting them into 
the context of generalized random cipher in the next section.

\section{\textbf{General quantum block cipher}}
Since block ciphers based on quantum data locking do not have operational meaning, we discuss here the construction of 
general quantum block ciphers based on Yuen's framework.

\subsection{\textbf{Structure of mathematical block cipher}}
Block ciphers are a method of dividing plaintext into blocks of length $n$ bits, and generating ciphertext of 
the same length using a secret key $K$ and a scramble circuit. Typical examples are DES (Data Encryption Standard) 
and AES (Advanced Encryption Standard).
Because the secret key is fixed, some ingenuities for encryption of long plaintext $|X|=2^{|K|}$bits input at 10Gbit/sec 
are required for the scramble circuit. The iteration of the algorithm for each plaintext block is called a mode, 
and a typical example CBC(cipher block chaining) mode.

The security depends on the complexity of this scramble structure.
In general, ciphertext-only attacks, known plaintext attacks, and various other attacks are set. 
Thus, security must be guaranteed against each of them under several mathematical assumptions.

\begin{figure}
\centering{\includegraphics[width=7cm]{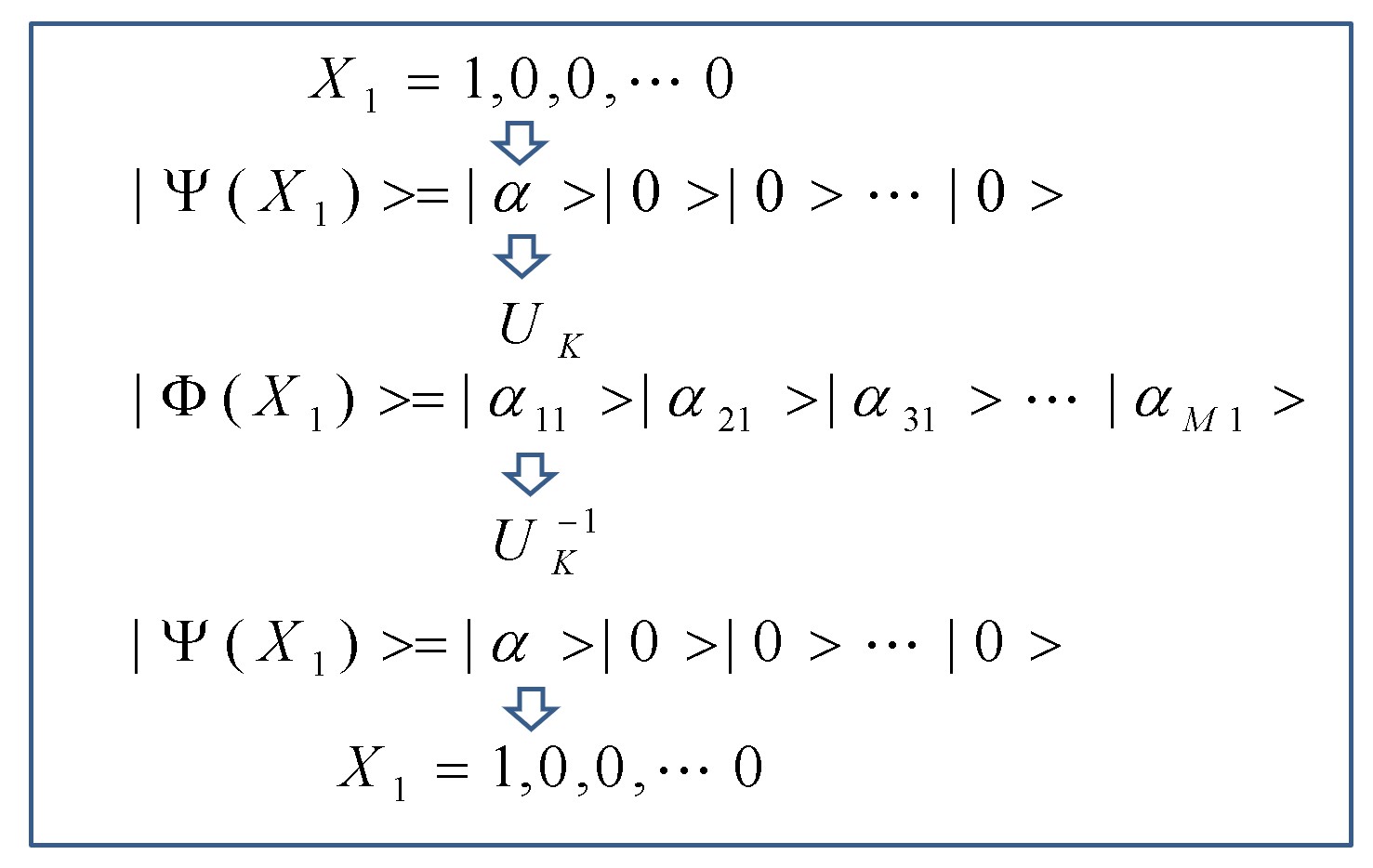}}
\caption{Structure of quantum block cipher based on coherent pulse position modulation CPPM.}
\end{figure}

\subsection{\textbf{Quantum block cipher form based on PPM}}
In general, if we follow the structure of the conventional block cipher, plaintext of length $n$ bits 
is treated as one block code and the number of corresponding codeword is generally $M=2^n$.
It is difficult to directly map these to quantum states of the same length $n$ with security guarantee.

Lloyd [16] and Guha [17] adopt Yuen's coherent pulse position modulation (CPPM) [18] to consider block ciphers 
based on quantum data locking.
Since CPPM itself already has the functionality of generalized quantum symmetric key cipher, no new functionality 
can be found by considering it as a quantum data locking method.
Here, we will proceed with the discussion as the construction principle of a block cipher using the PPM method.

Fig.6 shows its structure. The plaintexts of $M$ are mapped to PPM(pulse position modulation) code of $M$ slots, 
and these are constructed by quantum states as follows.
\begin{eqnarray}
S_1: |\Psi(X_1) > &=&|\alpha>_1|0 >_2 |0 >_3 \dots |0 >_M \nonumber \\
S_2 :|\Psi(X_2) >&=&|0 >_1|\alpha >_2 |0 >_3 \dots |0 >_M \nonumber \\
&&\vdots  \\
S_M:|\Psi(X_M) > &=&|0 >_1|0 >_2 |0 >_3 \dots |\alpha >_M \nonumber
\end{eqnarray}
These are to be regarded as a codeword set $\{S_m\}, m=1,2,\dots, M$.

Next, when a $2M$-dimensional unitary transformation ${\bf U}$ is applied to an optical pulse signal train
 with the above quantum state vector sequence of Eq(27), the amplitude $\alpha:(\theta=0)$ localized at one place can be 
distributed to all slots. Then, the quantum state vector sequence of the pulse train automatically becomes
 one of the following signal set. 
\begin{eqnarray}
S_1^T:|\Phi_1 > &=&|\alpha_{11}>|\alpha_{21}> |\alpha_{31} > \dots |\alpha_{M1} > \nonumber \\
S_2^T:|\Phi_2 > &=&|\alpha_{12}>|\alpha_{22} > |\alpha_{32}> \dots |\alpha_{M2}> \nonumber \\
&&\vdots \\
S_M^T:|\Phi_M > &=&|\alpha_{1M}>|\alpha_{2M} >|\alpha_{3M}> \dots |\alpha_{MM} > \nonumber
\end{eqnarray}
where $\{\alpha_{l,m}\}$ is complex amplitude. These $\{S^T_m\}$ are called random code word.
This transformation corresponds to quantum block cipher scheme.
The theory of unitary transformations which implement such transformations is explained in the Appendix.

Then, in order to avoide the repeated key for the long plaintext, 
such transformations are controlled by a running key sequence $K^R$ from  a pseudo-random number generator (PRNG) 
with a short secret key $K$.
Thus, one of the above set is randomly sent out into a communication channel as a ciphertext.

Here, let ${\bf U}_K=\{U(k^R_m)\}$, $m =1,2,3, \dots, M$ be the set of unitary transformation for 
this PPM modulation signal sequence.
The output runing key sequence from the pseudo-random number generator with a secret key $K$ 
is divided into $log M$-valued running key sequence $k^R$.
The running key sequence determines which unitary transformation is to be acted upon.
 When $U(k^R_m)$ is selected, if the input is from $|\Psi (X_1)>$ to $|\Psi (X_M)>$,
the resulting output quantum state sequences are as follows.
\begin{eqnarray}
|\Phi_1(k^R_m,X_1) >&=& U(k^R_m)|\Psi (X_1)>\nonumber \\
|\Phi_2(k^R_m,X_2) >&=&U(k^R_m)|\Psi (X_2)>\nonumber \\
&& \quad \quad \vdots \\
|\Phi_M(k^R_m,X_M) >&=& U(k^R_m)|\Psi (X_M)> \nonumber 
\end{eqnarray}

Bob's receiver uses the information of the unitary transformation.
He transforms the random code word Eq(29) by the inverse unitary transformation to convert back to the original signal of Eq(27).
As a result, he can discriminate them by photon counters almost without error at each slot as follows:
\begin{equation}
P(0|0)=1, \quad P(0|\alpha)=\exp \{-|\alpha|^2\}
\end{equation}
When $|\alpha |^2 \gg 1$ and $M \gg 1$, the average error probability for the code word is given as follows:
\begin{equation}
{\bar P^B_e}=(1-\frac{1}{M})\exp \{-|\alpha|^2\}\cong \exp \{-|\alpha|^2\} \ll 1
\end{equation}

On the other hand, since the eavesdropper does not know which unitary transformation is used in the modulator, 
it is necessary to directly receive optical signals of the random quantum code word of Eq(29), and to identify the vast number of
 possible combinations of complex amplitudes by quantum measurement.

\subsection{\textbf{Security issue}}
\subsubsection{\textbf{Ciphertext only attack on data}}
We first consider a ciphertext-only attack on the data (plaintext).
At this time, the task of the eavesdropper is to recover the $n$-bit plaintext information from the group of 
signals converted into $M$-ary random codes.
Error analysis for this model is performed using the theory of Gallager et al (Fig.7).

Eve's strategy is to receive the $M$-ary random code as accurately as possible.
Without explicitly specifying the unitary transformation, Yuen has shown that the lower bound of 
their average error probability is as follows [18]:
\begin{eqnarray}
{\bar P}_e^E &>& 1-\left \{\Phi(z)\right \}^L \Phi(z-2S) \\
 \Phi(z)&=& \frac{1}{\sqrt 2\pi}\int^z_{-\infty} exp\{-t^2/2\}dt \nonumber \\
L&=&\log M, \quad S=|\alpha|_{PPM}^2 \nonumber
\end{eqnarray}

Thus, the security is guaranteed by the difference between the error probabilities for Bob's receiver and Eve's receiver 
 tapping into the communication channel.
That is, the average error probability for code words upon reception of a signal system with quantum state vector sequences of
 Eq(27) and Eq(29) is overwhelmingly large for the latter, while that for the former is almost zero.

Consequently, when $|\alpha|^2 \gg1$, the eavesdropper's error
 becomes ${\bar P}_e^E \longrightarrow 1$  when $M$ is exponentially large. Then one has the following relation.
 \begin{equation}
 H(X|Y) \cong H(X), \quad under |K| \ll |X|
 \end{equation}

\begin{figure}
\centering{\includegraphics[width=7cm]{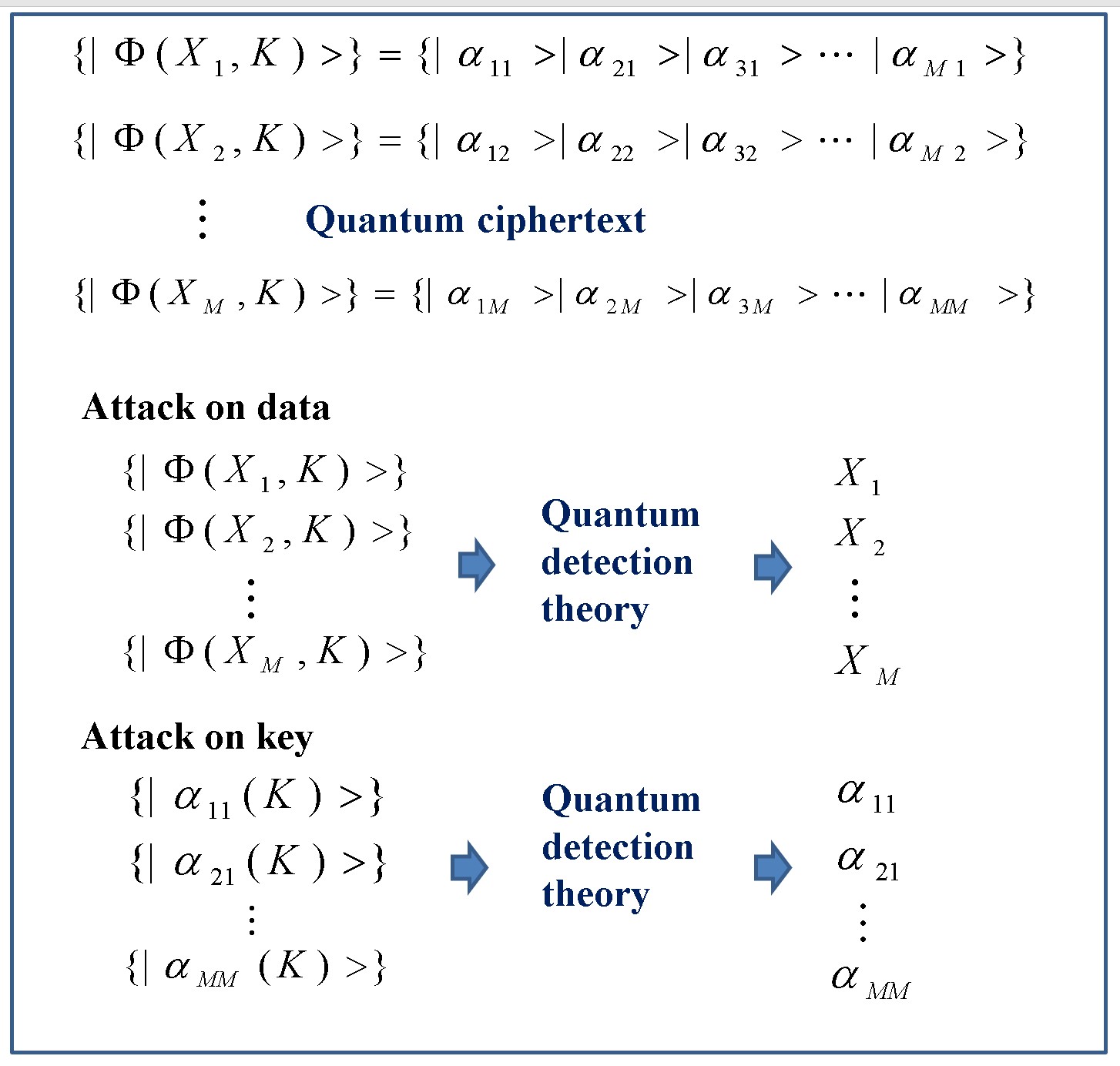}}
\caption{Structure of quantum measurement processes of attack on data and key. 
On data, the block information is the target and the set of element state is the target for key.}
\end{figure}

\subsubsection{\textbf{Attack on key}}
Like quantum stream ciphers, it uses pseudorandom numbers. Therefore one requires analysis of attacks against the secret key.
A key attack corresponds to receiving as accurately as possible the running key information from the quantum ciphertext.
A quantitative evaluation is not possible unless the unitary transformation is given concretely.
Here we will only explain the logic of the attack process.

When estimating the plaintext, it was a maximum likelihood estimate of the codeword based on the $M$-slot measurements.
The running key information is obtained by individually estimating the $M$-slot coherent state signal, 
which is a unitary transformed PPM signal.
In general, the signal in each slot is multi-valued or continuous, so the error becomes large. 
Furthermore, the success probability of the entire $M$ slots becomes very small, so the probability of 
determining the running key determined by the signal value of the $M$ slots approaches almost zero.
From the above, when only the ciphertext is used, $H(K|Y)$ is evaluated, and when the plaintext is known, 
$H(K|Y,X)$ is evaluated, based on the signal error probability.

\subsection{\textbf{Systemic flaw}}
Difficulties of two issues remain when trying to realize the above scheme. 

\textbf{(a)} One is that it is difficult to realize a device that performs unitary transformations in the on-off based PPM, 
 because it involves the vacuum state $|0> $. This is one reason why no experimental attempts have existed to date.

\textbf{(b)} The other is that the length $M$ of Eq(27) must be made considerably large to achieve the high security as 
denoted in the above.
It means that the delay of the encryption (letency) becomes a major drawback.
When requiring error probabilities by the exponential speed of convergence and eliminating the delay characteristics, 
the following increase in the bandwidth is necessary.
\begin{equation}
W_{cppm} \approx M \times B_S
\end{equation}
where $M \rightarrow \infty$. It means that when the speed of convergence of error probabilities and 
the associated delay effects are taken into account, one needs exponential increase of $M$.
 The above quantity is extremely large as a baseband for receiver circuit.\\

In order to apply this technology to a practical optical communication system, 
we have to solve the above issues.

\section{\textbf{Frequency-phase PPM as block cipher}}

A new PPM method to overcome the above difficultiesis was proposed in literature [19] (Fig.8). 
In the new system, $M$ is the number of frequency mode, and $J$ is the number of phase signal at each frequiency.
Then the number of quantum ciphertext becomes $J^M$.
Here, we explain its security evaluation. For the detailed structure of the system, see reference [19].

Prepare set of $M$-ary PPM code signal based on binary PSK by $M$ frequency modes.
Alice converts original PPM signal based on binary PSK into random code (Eq(51)) in the appendix) as quantum ciphertext 
using the unitary map ${\bf U}_K$ driven by the running key sequence from PRNG (See appendix). 
Then these random codes are transmitted to the legitemate reciver.

\begin{figure}
\centering{\includegraphics[width=8cm]{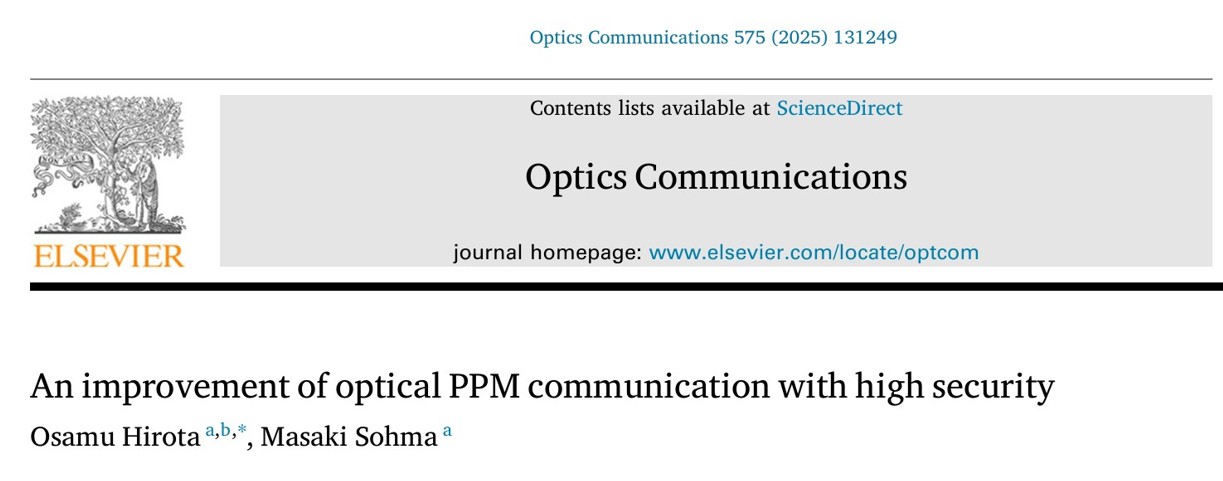}}
\caption{Reference [19] for Fm-phase PPM. Optics Communications,vol-575, 131249,2025. Open access.}
\end{figure}

\subsection{\textbf{Receiver system for legitemate reception}}
The legitimate receiver converts $M$-ary random code as quantum ciphertext into original $M$-ary  PPM signal based on binary PSK 
using the inverse unitary transformation  ${\bf U}^{-1}_K$ based on the same PRNG. 
The method for decoding binary PSK-PPM of $M$ frequency modes with strong light can be used  maximum likelihood decision or 
decision based on individual slot measurement. 
So, the error is very small and communication is not affected. Then he can obtain directly correct plaintext.

\subsection{\textbf{Receiver system for eavesdropper}}
The Figure 5 in the reference [19] shows the optimum receiver system for an eavesdropper when one adopts the proposed scheme.
Since the eavesdropper does not know the secret key,
 she cannot carry out the inverse unitary transformation on the received quantum ciphertext. 
The random codes as quantum ciphertext lose its decoding gain by random unitary transformation to the original phase PPM, 
so quantum optimal receiver with individual measurements is adopted.

The general properties of quantum Bayes rule and quantum minimax rule for the symmetric optical signals of multi parameter 
are given by Ban [20]. One of them is as follows:\\

{\bf Theorem-6} : \\
If the signal set $\{|\alpha_j>\}, j=\{1,2, \dots, J\}$ is a symmetric, the optimum quantum measurement 
for each signal ($\Pi_j$) is given by using Gram operator $H$ as follows:
\begin{eqnarray}
&&\Pi_j =|\mu_j><\mu_j|  \\
&&|\mu_j>= H^{-1/2}|\alpha_j> \nonumber \\
&&H= \sum_{j=1}^J |\alpha_j><\alpha_j| \nonumber
\end{eqnarray}
and the optimum  quantum Bayes solution is 
\begin{equation}
{\bar P}_e =1-|<\alpha_1|H^{-1/2}|\alpha_1>|^2
\end{equation}

Let us assume that the eavesdropper adopts a simple decoding procedure with the above formula. 
 By applying calculation methods for the above formula on PSK of Osaki [21] and Kato [22] and 
also the quantum minimax theorem [23] to our scheme, 
we have the Eve's error probability for $M$-ary random codes in this case as follows: 
\begin{eqnarray}
{\bar P}_e^E &\cong&1- \left \{\frac{1}{J^2}(\sum_{l=1}^{J} {\sqrt \lambda_l})^2 \right \}^M  \\
\lambda_l&=&\sum_{k=1}^{J} <\alpha_1|\alpha_k >u^{-(k-1)l} \nonumber
\end{eqnarray}
where $u=\exp[i(2\pi/J)]$. This requres a numerical analysis for the visualization. Fortunately in the case of $J \gg 1$ 
under the fixed $M$ frequency modes,
 a set of the quantum states may be treated as a quantum state system corresponding to a continuous signals. 
So the quantum optimum measurement can be approximated by Yuen-Lax quantum Cramer-Rao bound [24].
As a result, one can approximate the optimum measurement by a heterodyne measurement.
Thus, to demonstrate intuitive understanding, we can assume that Eve adopts a heterodyne measurement.
Then we here can use the conventional Bayes rule, assuming  a priori probability is $1/J$ for each mode.
The error performance can be given in the case of $J \gg 1$ and  $J^M \gg 1$ as follows:

The outputs of the heterodyne are the analog current denoted by 
$\{{\hat \theta}_{11}, {\hat \theta}_{22}, \dots, {\hat \theta}_{MM} \}$, 
in which each phase is estimated from complex amplitude.
The error probability of ciphertext in such a system based on PSK is given by 
\begin{equation}
{\bar P}_e^E \cong 1-\left \{erf(\frac{\Delta}{2\sigma_{he}})\right \}^M 
\end{equation}
where $\sigma_{he}=1$ is the quantum noise effect in heterodyne measurement.
$\Delta$ is the phase difference between neighboring phases in $J$-ary PSK.
\begin{eqnarray}
&& \Delta=|\alpha|(1-\cos\delta)^{1/2} \\
&& \delta =2\pi/J
\end{eqnarray}
Thus, in the case of our scheme, one can control the error performance by $J$ under the fixed $M$.
In other words, the security can be guaranteed with a small $M$.
So one can show that the average error probability for $M$-ary channel becomes as follows:
\begin{equation}
{\bar P}_e^E \sim 1, \quad J \gg 1, \quad M:fixed
\end{equation}
As a result, Eve's Shannon mutual information of quantum measurement channel to the $M$-ary random code is 
\begin{equation}
C_1 \longrightarrow 0, \quad J \gg 1
\end{equation}
Thus, even if the drive of the random selection is pseudorandom, its correlation properties are masked by quantum noise, 
and the unicity distance of secret key can be made extremely large based on Eq(10).
\begin{equation}
n^Q_1 \gg |K|
\end{equation}
Concequently, it is possible to realize quantum block ciphers that enable high-speed, long-distance transmission 
within a finite bandwidth.

\section{\textbf{Towards real world applications of quantum stream cipher with lifiting the Shannon impossibility theorem}}
\subsection{\textbf{Current status of experiment}}
Quantum stream cipher is a technology that provides the encryption service to the high-speed data communications. 
Conferences are being held frequently to discuss how to deploy this technology. The Fig.9 shows an example.
Examples of a case study for practical application are shown in Fig.10 and Fig.11.
The first example is a data encryption system to ensure the security of submarine cables (Fig.10), 
while the latter is to ensure the security of existing optical global network lines (Fig.11).
The preliminary experiment towards the application to submarine cable system was demonstrated [25]. 
In this case, authentication and key distribution are not necessary, so data communication is directly carried out by 
the quantum stream cipher transceiver.

In addition, the Tamagawa University group has successfully demonstrated an overall system of a novel quantum-enhanced secure 
link architecture for future global network applications (Fig. 10) using actual fiber optic lines embedded in a railway. 
The experimental demonstration was reported in [26], and the overall operation on 
the use of the post-quantum cryptography (PQC) was explained in [27].
 In the following, we introduce the operation scheme according to [27].

The experimental devices were placed in three locations: the Ebina Station of the Sagami Railway in the city of Ebina (Alice), 
the Yamato Station in the city of Yamato (Bob), and the Tamagawa University in the city of Machida. 
The server providing user authentication and key exchange services via an Internet connection was placed at Tamagawa University. 
Y-00 transceivers for secure data communication over optical fiber lines and client PCs for user authentication and key exchange 
were placed at the Ebina and Yamato stations, respectively. 

The experiment involves three steps. The first step is user authentication, where the server at Tamagawa confirms that the users 
(Alice and Bob) who want to connect are indeed legitimate users using PQC digital signature algorithms. The second step is 
to generate a secret key between the authenticated users using PQC key exchange algorithms. 
Finally, the Y-00 transceiver with the generated secret key is used to perform high-speed secure data communication over 
the embedded fiber optic lines. The specifications used in each step are summarized as follows:

\textbf{Authentication}: \\
In this step, they used three PQC digital signature algorithms; CRYSTALS-Dilithium, Falcon, and Rainbow. 
These three were the Round 3 Finalists and the first two are the selected algorithms of the NIST PQC project. 
In this experiment, the users (Alice and Bob) registered their public key (as a verification key) with the server prior 
to the experiment, and on the day of the experiment, they received user authentication services via the Internet using private key 
(as a signature key). To implement PQC algorithms, they used the libraries provided by the Open Quantum Safe Project.

\textbf{Key-establishment}: \\
In this step, they used CRYSTALS-KYBER, a PQC key-encapsulation mechanism, and SHA256, a hash function, where CRYSTALS-KYBER 
is one of the selected algorithms of the NIST PQC project. Alice and the server share a random sequence using CRYSTALS-KYBER. 
Bob and the server also share another random sequence in the same manner. The server announces the locations of 
the mismatched bits in the random sequences to both Alice and Bob. By performing the sifting operation, 
Alice and Bob share a pre-key which has an indefinite length. Applying SHA256 to the pre-key, 
Alice and Bob can share a 265-bit secret key.

\textbf{Secure data communication}: \\
We used an ISK-type quantum stream cipher transceiver that operates at 1 Gbit/s in the optical C band. 
The optical fiber cable connecting Ebina and Yamato is buried along the tracks, 
so it is subject to various physical changes such as temperature and vibration, but it operated normally.

\subsection{\textbf{Development of security-guaranteed systems beyond 100 Gbps}}
It is becoming essential for cryptography to keep up with the rapid development of optical communication technology. 
Our cryptography needs to keep up with the demands of the optical communication researchers.

The generalized quantum stream cipher with real-time support of 100 Gbps $\sim$ 10 Tbit/sec over a communication distance of 1000 km 
$\sim$ 10,000 Km will become the standard. 
Optical communication equipment manufacturers are developing high-speed DAC(Digital to analog converter) and 
ADC(Analog to digital converter) to put encryption devices into practical use. 
Development of network-compatible optical switches and optical amplifiers is also progressing smoothly. 
In addition, experiments on the product cipher mechanisms for security against KPA  have begun by Futami and his group [28]. 
Quantum block cipher is a future challenge, and verification experiments of Sohma theory shown in the appendix will be important.\\

\begin{figure}
\centering{\includegraphics[width=7cm]{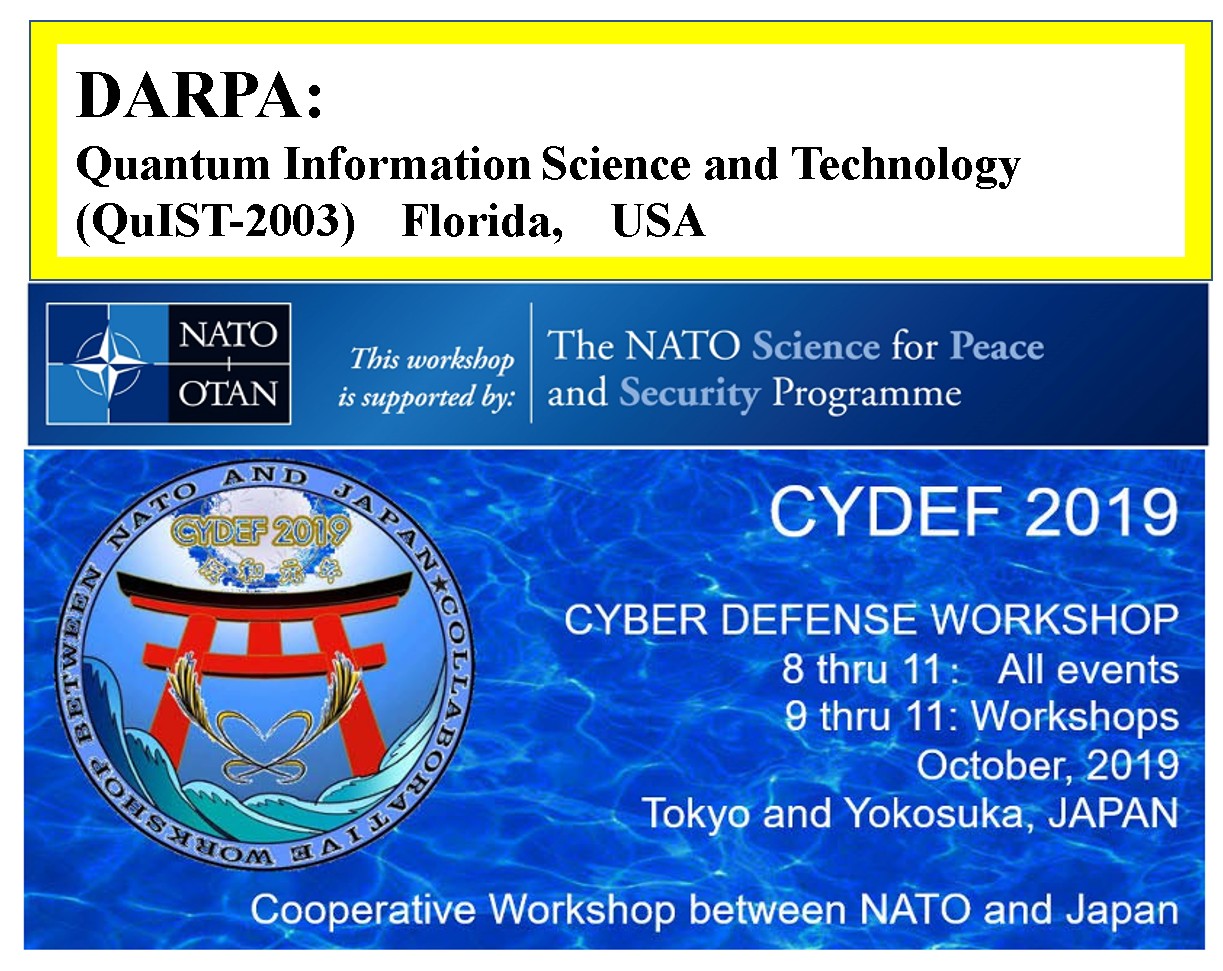}}
\caption{DARPA Workshop on quantum information science and technolpogy at 2003 and NATO workshop of CYDEF-2019.
A case study of quantum stream cipher was discussed.}
\end{figure}

\begin{figure}
\centering{\includegraphics[width=8cm]{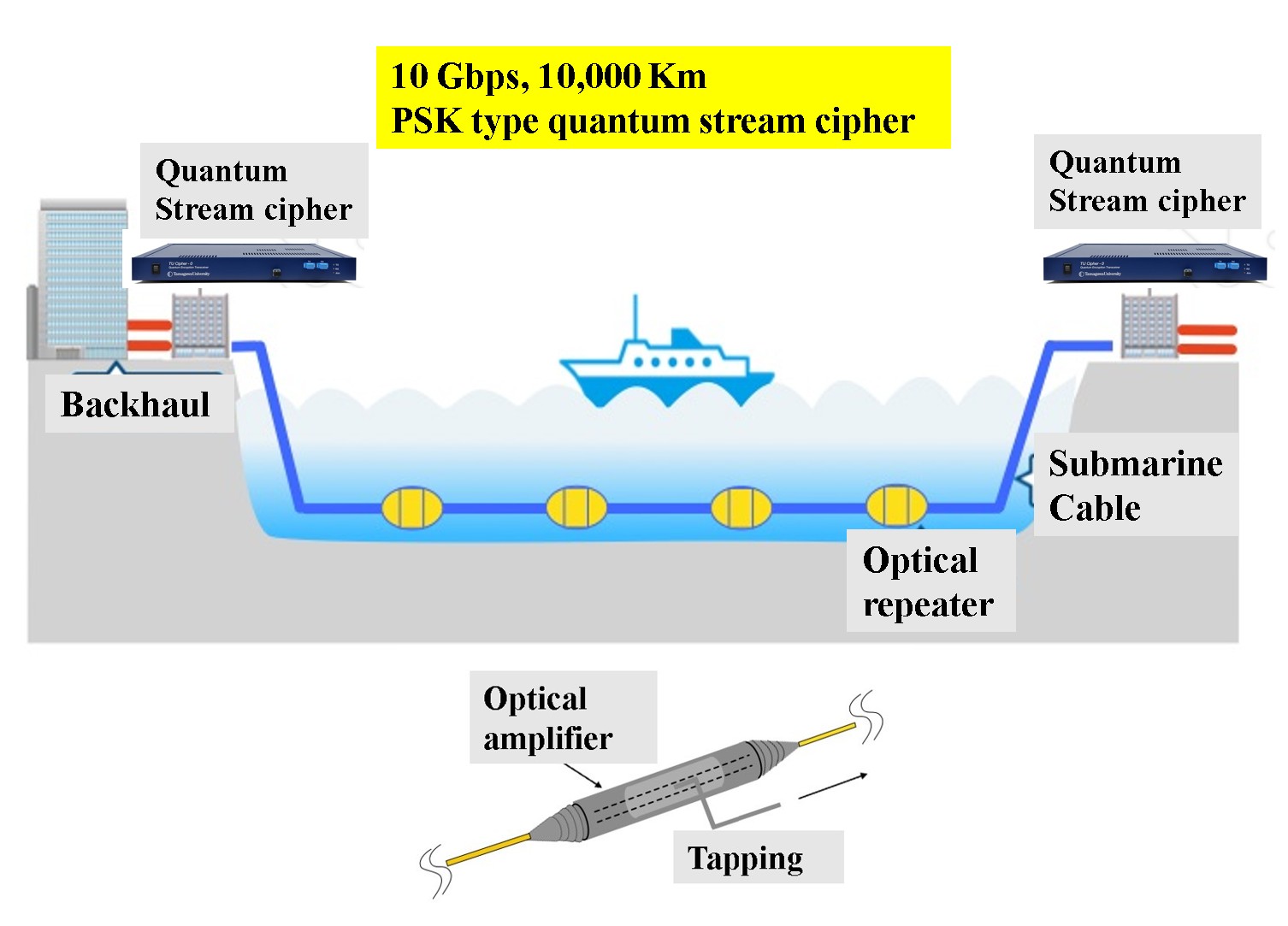}}
\caption{Final target image of application to submarine cable system with eavesdropping incidents in the amplifier repeater system.
Tanizawa and Futami have successfully conducted a preliminary transmission experiment by using a land cable of 10,000 Km with 1Gbit/sec 
quantum stream cipher tranceiver.}
\end{figure}

\begin{figure}
\centering{\includegraphics[width=8cm]{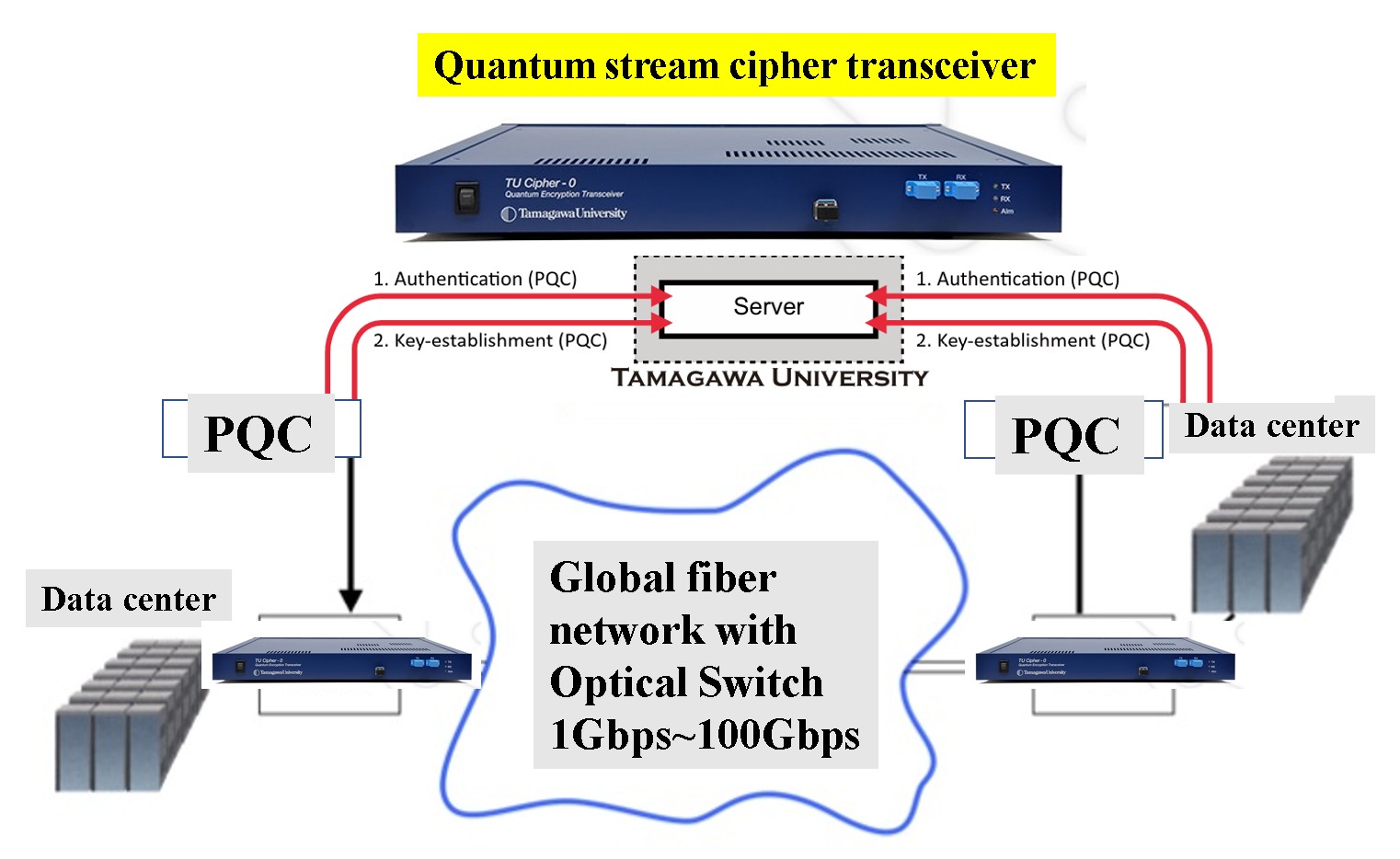}}
\caption{Final target image of global optical network application. Futami,Tanizawa, and Kato have successfully conducted 
a preliminary experiment 
on the total scheme with PQC for authentification and key establishment at the Railway Optical Communications System.}
\end{figure}

\section{\textbf{Conclusion}}

This paper has provided a detailed explanation on the lifting the Shannon impossibility theorem discussed in the previous paper. 
These will be appropriate for an intuitive understanding on the reason of the lifiting the Shannon impossibility theorem.
It is important to realize that although it is about ciphertext-only attacks on the data, the essence extends to 
known-plaintext attacks on the key [2,3].

In addition, we have explained that quantum stream cipher and quantum data locking cipher as a block cipher have 
the same origin, but their encryption schemes are different [16,17,29,30].
But the latter requires engineering improvements. 
Then, we have provided the technical method for the improvement of the quantum data locking cipher as the block cipher.

The quantum stream cipher and quantum block cipher, which are the technologies that realize the generalized random cipher 
introduced here, can be used on real communication lines and do not have the inconvenience of having the same number of 
keys as the plaintext, as in one time pad. 
By simply sharing a 256-bit secret key, it is now possible to encrypt data of more than 10 Gbit/sec
in an information-theoretically secure manner on data (Fig.12).

On the other hand, a great deal of implementation experiments [31,32,33,34] have already been carried out on 
the standard quantum stream cipher, and a project is also being planned for commercialization. 
In addition, experimental challenges have begun on generalized quantum stream ciphers with ultra-high performance ranging 
from 10 Gbps (10,000 km) to 100 Gbps (1,000 km). For summary, see American Physical Society TV [35]. 
We will report any further results if any.

\begin{figure}
\centering{\includegraphics[width=8cm]{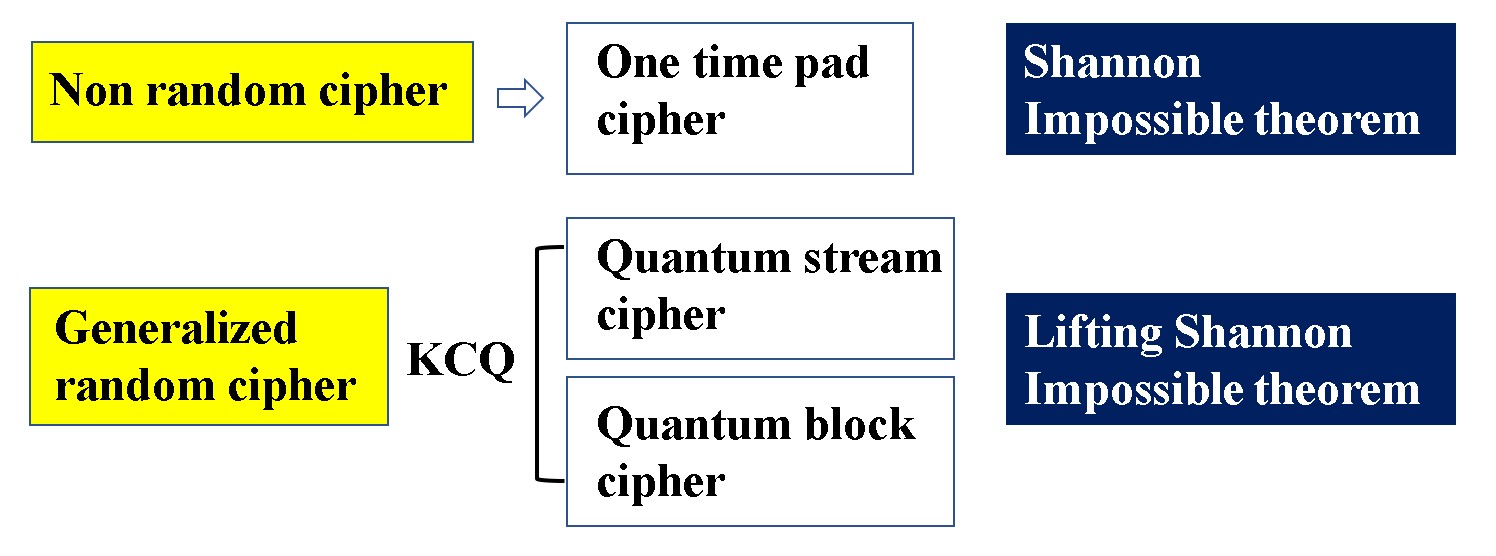}}
\caption{The security on data of OTP and KCQ is the same. In the conventional cryptology, one time pad cipher is 
the final technology constrained by the Shannon impossibility theorem. 
In the KCQ, security on key is neccesary and the generalization of 
the standard type is required [2,3]}
\end{figure}

\section*{\textbf{Appendix}}
\section*{\textbf{General theory of encryption by unitary transformation}}
Here we introduce the encryption theory using unitary transformation by Sohma [36,37].

\subsection{\textbf{Quantum Gaussian state:}}
We can define an operator ${\bf V}$ (Eq(46)) that extends the Heisenberg commutation relation for self-adjoint operators on 
Hilbert spaces to the Weyl-Segal commutation relation. We use it to define quantum characteristic functions as follows.

 First, let us consider a general $M-$ary coherent state (Gaussian state):
\begin{equation}
|\Phi >=|\alpha_1>|\alpha_2>|\alpha_3> \dots |\alpha_N>
\end{equation}
From Stone-von Neumann theorem [38], the quantum characteristic function for the class of quantum Gaussian state is given as follows:
\begin{eqnarray}
\Phi ({\bf z})&=&Tr {\bf U}|\phi><\phi| {\bf U}^{\dagger} {\bf V}({\bf z}) \nonumber \\
&=&Tr |\phi><\phi|{\bf V}({\cal L}^T {\bf z}) 
\end{eqnarray}
where
\begin{eqnarray}
{\bf V}({\bf z})&=&\exp \{i {\bf R}^T {\bf z}\} \\
{\bf R}&=&[({\bf q}_1,{\bf p}_1), \dots ,({\bf q}_M,{\bf p}_M), ]^T 
\end{eqnarray}
and where $({\bf q}_i,{\bf p}_i)$ are the canonical conjugate operators. Then ${\cal L}$ is a symplectic matrix, and it is given by 
\begin{eqnarray}
&&{\cal  L}=\left (
\begin{array}{ccccc}
r_{11}e^{i\theta_{11}}& \dots & \dots &r_{1M}e^{i\theta_{1M}}   \\
r_{21}e^{i\theta_{21}}& \dots & \dots & r_{2M}e^{i\theta_{2M}}\\
\vdots & \dots & \dots & \vdots \\
r_{M1}e^{i\theta_{M1}} & \dots & \dots  & r_{MM}e^{i\theta_{MM}}  \\
\end{array}
\right )
\end{eqnarray}
We call ${\bf U}$ the unitary operator associated with symplectic transformation ${\cal L}$.

\subsection{\textbf{Randomization as encryption:}}

A set of ${\cal L}$ with different elements is prepared. One ${\cal L}$ is selected from the set using a pseudorandom sequence 
with the secret key $K$ 
A quantum ciphertext is generated by a unitary transformation associated with the selected ${\cal L}$ as follows:

Here let us denote a vector of complex amplitudes $\alpha$ of coherent state as follows:
\begin{equation}
\vec{\alpha}_{in}=(\alpha_1, \alpha_2, \dots , \alpha_M)
\end{equation}
then we have the following relation.
\begin{equation}
\vec{\alpha}_{out}={\cal L} \vec{\alpha}_{in}=(\alpha_1^{out}, \alpha_2^{out}, \dots , \alpha_M^{out})
\end{equation}
As a result, the unitary transformation for the coherent state sequence is given  as follows:
\begin{equation}
{\bf U}|\Phi> =|\Phi_{out}> =|\alpha_{1}^{out}>|\alpha_{2}^{out}> \dots |\alpha_{M}^{out}>
\end{equation}
Thus, randomization of complex amplitudes through ${\cal L}$ converts the basic codes of coherent states into quantum ciphertext with 
arbitrary complex amplitude by ${\bf U}$.
When $r_{i,j}=1, \forall i,j$, the above is called phase randomization.

On the other hand, Bob's decryption procedure involves applying the inverse ${\bf U}^{-1}$ of a pseudorandomly 
selected unitary transformation to the quantum ciphertext using the same pseudorandom numbers.
As a result, a receiver with the key can always receive the basic quantum state code signal before the randomization.
For a more detailed discussion, see the literature [37,38].
 
 \begin{figure}
\centering{\includegraphics[width=8cm]{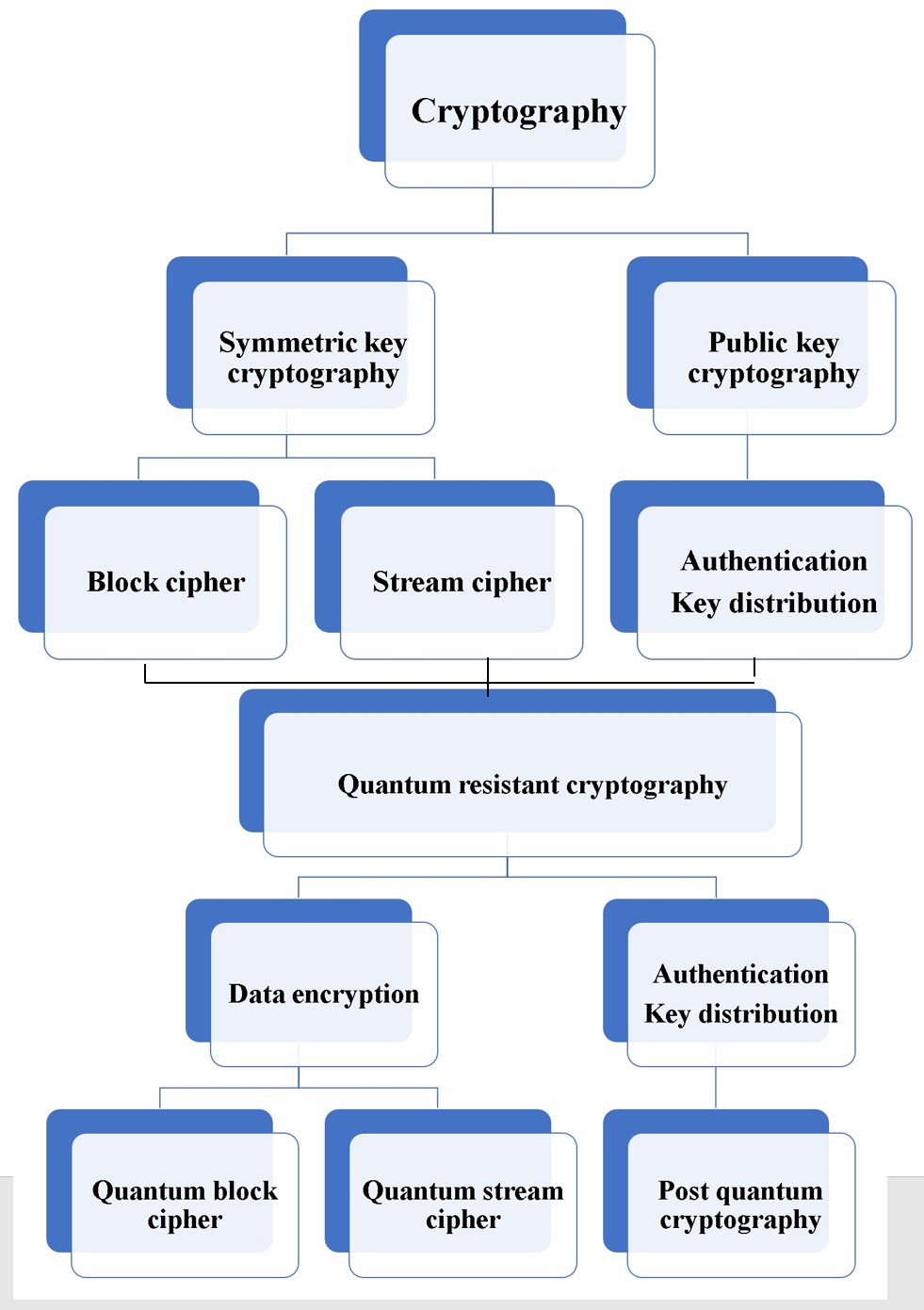}}
\caption{Yuen's concept for the next generation technology in secure communication networks construted by 100 Gbit/sec 
$\sim$ 10 Tbit/sec speed and global optical netwok including 10,000Km long submarine system.}
\end{figure}

 \section*{\textbf{Acknowledgements}}
I am grateful to K.Kurosawa, M.Sohma, and K.Kato for the discussions I had with them. I would like to express my gratitude to  
F.Futami, T.Usuda, K.Nakahira and K.Tanizawa for their activities. Finally, we would like to ask for the cooperation of 
all relevant parties in realizing the final form of the next-generation communications network that Yuen envisions (Fig.13).

\end{document}